\newcommand{\wt}[1]{\widetilde{#1}}
\newcommand{\wh}[1]{\widehat{#1}}
\newcommand{\ket}[1]{\left| #1 \right\rangle}

\def\hs{\mathcal{H}^\text{comp}}
\def\junk{irrelevant }
\def\hsjunk{\mathcal{H}^\text{irr}}
\def\ehs{\mathcal{H}^\text{phys}}
\def\ham{H}
\def\op{\mathcal{O}}
\def\id{\mathds{1}}

\def\mk{$k$}
\def\Hloc{H^*}
\def\Hwrong{H^\text{irr}}
\def\psit{\wt{\psi}}
\def\hamt{\wt{\ham}}
\def\ut{\mathcal{U}}
\def\fB{\mathcal{B}}
\def\fQ{\mathcal{Q}}
\def\At{{\wt{A}}}

\def\stdham{H_{\text{std}}}
\def\nnloc{N_{nl}}
\def\nloc{N_l}
\def\ntot{N_{tot}}
\def\Kt{\wt{K}}
\def\comm{\text{comm}}
\def\anti{\text{anti}}
\def\target{computational }
\def\physical{physical }
\documentclass[superscriptaddress,aps,pre,reprint,nofootinbib]{revtex4-1}
\usepackage{amsmath}
\usepackage{braket}
\usepackage{dsfont}
\usepackage[final]{graphicx}
\usepackage[usenames, dvipsnames]{color}

\begin{document}

\title{Simulating highly nonlocal Hamiltonians with less nonlocal Hamiltonians}
\author{Y. Suba\c{s}\i} 
\affiliation{Department of Chemistry and Biochemistry , University of Maryland, College Park, MD 20742, U.S.A.}
\author{C. Jarzynski}
\affiliation{Department of Chemistry and Biochemistry , University of Maryland, College Park, MD 20742, U.S.A.}
\affiliation{Institute for Physical Science and Technology, University of Maryland, College Park, MD 20742, U.S.A.}
\begin{abstract}
The need for Hamiltonians with many-body interactions arises in various applications of quantum computing.
However, interactions beyond two-body are difficult to realize experimentally.
Perturbative gadgets were introduced to obtain arbitrary many-body effective interactions using Hamiltonians with two-body interactions only.
Although valid for arbitrary $k$-body interactions, their use is limited to small $k$ because the strength of interaction is $k$'th order in perturbation theory.
In this paper we develop a nonperturbative technique for obtaining effective $k$-body interactions using Hamiltonians consisting of at most $l$-body interactions with $l<k$.
This technique works best for Hamiltonians with a few interactions with very large $k$ and can be used together with perturbative gadgets to embed Hamiltonians of considerable complexity in proper subspaces of two-local Hamiltonians.
We describe how our technique can be implemented in a hybrid (gate-based and adiabatic) as well as solely adiabatic quantum computing scheme.
\end{abstract}

\maketitle

\section{Introduction}

Many-body interactions are in general difficult to realize in physical systems used in quantum information processing. 
However, there is substantial theoretical interest in Hamiltonians with such interactions. 
One motivation to study them comes from a question in complexity theory: which Hamiltonians are capable of universal adiabatic quantum computation (AQC)? 
Kitaev~\cite{Kitaev2002Book} showed that the ground state energy problem of the 5-local Hamiltonian is quantum-Merlin-Arthur (QMA)-complete. 
This result was later strengthened by Kempe and Regev~\cite{Kempe2003QIC} showing the same for the 3-local Hamiltonian problem. 
Finally, by introducing ``perturbative gadgets" to construct a 2-local Hamiltonian whose low energy effective Hamiltonian approximates a given 3-local Hamiltonian, Kempe, Kitaev and Regev~\cite{Kempe2006SIAM} showed that the 2-local Hamiltonian problem is QMA-complete as well.  
In order to strengthen these results it is desirable to replace the perturbative gadgets with nonperturbative techniques.

On the more practical side, \mk-local Hamiltonians are necessary to tackle difficult optimization problems like $K$-SAT.  
These problems are also valuable in order to test the power of adiabatic quantum algorithms~\cite{Farhi2002arXiv}, because they are classically more challenging than those that can be directly represented with 2-local Hamiltonians only, i.e., without any indirect embedding. 

Another area where the need for many-body interactions arises is the application of adiabatic quantum algorithm to quantum chemistry. 
Somma et al.~\cite{Somma2002PRA} has shown that fermions can be efficiently simulated using a quantum computer made of qubits only. 
In their scheme, the Jordan-Wigner transformation is used to represent the fermionic creation and annihilation operators in terms of the qubit Pauli operators. 
However, this transformation maps 2-body interactions between fermions into many-body interactions between qubits of any order. 
Bravyi and Kitaev~\cite{Bravyi2002AnnPhys} improved this scheme by finding a different transformation that produces interactions between qubits, the number of which scales only logarithmically in the system size.
Still, for sufficiently large systems, it is challenging to reduce these nonlocal Hamiltonians to 2-local Hamiltonians using perturbative gadgets.

Finally, we note that many-body interactions are necessary to implement some error correction schemes designed for adiabatic quantum computation~\cite{Jordan2006PRA} and quantum annealing~\cite{Young2013PRX, Pudenz2014Nat}. 
The basic idea is to encode each logical qubit using $k$ physical qubits in such a way that certain types of errors can be suppressed and/or corrected. 
An unavoidable consequence of the encoding is that some logical qubit operators are mapped to $k$-local operators, the implementation of which requires many-body interactions between the physical qubits.

In this paper we develop a nonperturbative technique to generate effective many-body interactions using Hamiltonians with fewer-body interactions. 
Our nonperturbative technique differs from the perturbative gadgets in several other aspects. 
First, unlike perturbative gadgets, each \mk-body interaction term requires the addition of a single ancillary qubit as opposed to \mk\, qubits. Second, the target Hamiltonian is not necessarily embedded in the low energy subspace of the physical Hamiltonian.
Finally, the technique described in this paper is not guaranteed to reduce the locality of an arbitrary Hamiltonian. 
It works best for Hamiltonians with few many-body interactions involving arbitrary number of qubits. 

The paper is organized as follows. 
In Sec.~\ref{sec:def} we establish the notation and state in detail the problem we address. 
In Sec.~\ref{sec:derivation} we present the derivation of the general theory. 
Special applications to AQC are presented in Sec.~\ref{sec:AQCapplication}. 
Sec.~\ref{sec:multiple} describes how multiple many-body interactions can be handled. 
The important question of how this technique can be implemented is discussed in Sec.~\ref{sec:implementation}. 
We conclude with brief remarks in Sec.~\ref{sec:conclusion}.

\section{Definitions and Conventions}
\label{sec:def}

Let $\hs$ represent the Hilbert space of the quantum system we are interested in, which consists of $N$ qubits. 
We will refer to this as the computational Hilbert space. 
A convenient basis for states in this Hilbert space can be constructed using tensor product of single qubit basis states
\begin{align}
\label{opbasis}
\ket{n} \equiv \ket{n_1} \otimes \ket{n_2} \otimes \dots \otimes \ket{n_N},
\end{align}
where $n = (n_1,n_2,\dots,n_N)$ and $n_i \in \{ 0,1 \}$. 
Any state in $\hs$ can be represented as:
\begin{align}
\label{wavefunction}
\ket{\psi(t)} &= \sum_{n \in \fB_N} c_n(t) \ket{n},
\end{align}
where $\fB_N$ is the space of $N$ binary numbers. 
The tensor product of single qubit Pauli operators is a convenient basis for the space of Hermitian operators in $\hs$:
\begin{align}
\label{def:op}
\wh{\op}^A \equiv \wh{\sigma}_1^{A_1}\otimes \wh{\sigma}_2^{A_2} \otimes \dots \otimes \wh{\sigma}_N^{A_N},
\end{align}
where $A = (A_1,A_2,\dots,A_N)$ and $A_i \in \{ 0,x,y,z\}$. 
By convention $\wh{\sigma}^0 \equiv \wh{\id}$.
Note that $\widehat{\op}^A$ are both Hermitian and unitary.
The Hamiltonian of the system of interest, i.e. the \target Hamiltonian,  can be expressed as: 
\begin{align}
\label{hamdecomp}
\wh{\ham}(t) = \sum_{A \in \fQ_N} \wh{H}^{A}(t) \equiv \sum_{A \in \fQ_N} h_A(t) \wh{\op}^A,
\end{align} 
where $\fQ_N$ is the space of N variables that can take the values $0,x,y,z$. 
In the rest of the paper we will drop the hats on the operators and suppress the time dependence of variables for brevity of notation. 

An operator is \mk-local if it acts non-trivially on at most \mk\, qubits. In terms of the basis operators $\op_A$ this amounts to having at most \mk\,  nonzero elements in the set $A$. 
To demonstrate this by an example let $A = (x,y,z,0,0,\dots,0)$. Then $\op^A = \sigma_1^x \otimes \sigma_2^y \otimes \sigma_3^z \otimes \id_4\otimes \dots \otimes \id_N$ is $3$-local. 
A Hamiltonian that is a sum of many terms is said to be $k$-local if each term in the sum acts on at most $k$ qubits. 
A single term of the Hamiltonian that acts on $k$ qubits will be referred to as a $k$-body interaction.

In practice, $1$-local terms are the simplest to realize in the laboratory. These are sometimes referred to as local ``fields". 
$2$-local Hamiltonians can also be realized experimentally, albeit with relatively more effort. They are sometimes referred to as ``interactions". 
However, it is quite a challenge to engineer $k$-local interactions for $k>2$. 

Most efforts in this direction involve embedding the \target Hamiltonian in the low energy sector of another Hamiltonian living in a larger Hilbert space. 
``Perturbative Gadgets" \cite{Kempe2006SIAM} are very useful in this regard. 
For a $k$-body interaction they require $k$ ancilla qubits. 
However, their use is limited to small $k$ because the nonlocal interaction emerges at the $k$'th order in perturbation theory~\cite{Jordan2008PRA}. 

A nonperturbative embedding of k-body interactions into 2-local Hamiltonians has been developed for the special case when all terms in the Hamiltonian share the same basis~\cite{Biamonte2008PRA}.   
In this manuscript, we describe a different nonperturbative technique which can be applied to any arbitrary Hamiltonian, albeit with varying success. 
In the rest of the paper we consider a \target Hamiltonian with a single many-body interaction term singled out:
\begin{align}
\label{originalH}
H = \sum_{A\ne \chi} \ham^A + \ham^\chi \equiv \Hloc + \ham^\chi\, , 
\end{align} 
where $\ham^\chi = h_\chi \op^\chi$ is \mk-local.  
The technique developed here will be most useful whenever $\ham^\chi$ is a $k$-body interaction that is difficult to realize experimentally and $\Hloc$ is a $2$-local Hamiltonian.
However, the technique is applicable to any Hamiltonian and the split in Eq.~\eqref{originalH} can be entirely arbitrary. 
For example, $\ham^\chi$ does not need to be the most nonlocal term in the Hamiltonian and there can be multiple many-body interactions.

We ask the following question: is there another Hamiltonian $\hamt$, possibly living in a larger Hilbert space, with a proper subspace in which it is identical to $\ham$? 
We refer to $\hamt$ as the \physical Hamiltonian. 
Note that we do not require $\Hloc$ to be $2$-local as $\ham$ might have multiple many-body interactions. 
Our goal is to find a different system the dynamics of which is simply related to that of the \target system and in which the \mk-local term $\ham^\chi$ is replaced by a less nonlocal interaction.

\section{Derivation}
\label{sec:derivation}

\subsection{Physical vs \target Hilbert Space}

We enlarge the Hilbert space of $N$ qubits by adding an ancilla qubit to obtain $\ehs = \mathcal{H}^\text{anc} \otimes \mathcal{H}^N$. 
The goal is to design the physical Hamiltonian $\hamt$ such that the physical Hilbert space splits into two subspaces, i.e. $\ehs = \hs \oplus \hsjunk$, such that within $\hs$ the dynamics evolves according to the \target Hamiltonian $H$ we desire to implement.~\footnote{For consistency of notation we should have used $H^\text{comp}$ for the \target Hamiltonian and $H^\text{phys}$ for the physical Hamiltonian, however, in order to avoid cumbersome notation we opted for $H$ and $\hamt$ instead.}
The other subspace $\hsjunk$ will be referred to as the \junk Hilbert space, because the dynamics there is not generated by the \target Hamiltonian. 
The dynamics in the \physical Hilbert space is governed by the Hamiltonian $\hamt$, which we wish to determine and in which the nonlocal term $\ham^\chi$ proportional to $\op^\chi$ will be replaced with a less nonlocal term. 

A convenient basis for the states in the \physical Hilbert space is 
\begin{align}
\label{stdbasis}
\ket{ \tilde{n} } \equiv\ket{n_0}\otimes \ket{n} = \ket{n_0} \otimes \ket{n_1} \otimes \ket{n_2} \dots \otimes \ket{n_N},
\end{align}
where the first entry is dedicated to the ancilla qubit. 
Any state in the \physical Hilbert space can be written as:
\begin{align}
\ket{\psit(t)} &= \sum_{\tilde{n} \in \fB_{N+1}} c_{\tilde{n}}(t) \ket{\tilde{n}}.
\end{align}
The physical Hilbert space is twice the size of the \target Hilbert space that we wish to simulate. 
We embed the dynamics in a subspace of the enlarged Hilbert space whose dimension matches that of the original N-qubit Hilbert space. 
In order to describe this subspace we need the following definitions:
\begin{align}
\label{basis}
\ket{n_{\pm}} &\equiv \ut \left(\frac{\ket{0} \pm \ket{1}}{\sqrt{2}}\otimes \ket{n}\right) \equiv \ut \left( \ket{\pm} \otimes \ket{n} \right),
\end{align}
where $\ut$ is a unitary operator (to be determined later) effecting a change of basis and $\ket{\pm}\equiv(\ket{0}\pm \ket{1})/\sqrt{2}$ are the eigenvectors of the $\sigma_x$ operator.  
$\ket{n_\pm}$ form a complete basis for the N+1 qubit system. 
More precisely, $\ket{n_+}$ form a complete basis on $\hs$ and $\ket{n_-}$ form a complete basis on $\hsjunk$. 
Using the definition \eqref{basis} we define projectors to two subspaces of the \physical Hilbert space
\begin{align}
P_\pm &= \sum_{n \in \fB_N} \ket{n_\pm}\bra{n_\pm}
\end{align}
These projectors satisfy 
$P_+ + P_- = \id_{N+1}$ and 
$P_+ P_- = P_- P_+ = 0$.
We want to embed the original dynamics of N qubits governed by the Hamiltonian \eqref{originalH} within the subspace $\hs$.~\footnote{Note that this strategy is quite different from the one used in perturbative gadgets where the embedding is done to the low energy sector of the theory.
In the approach described here energy does not play any role in the determination of the subspace into which the dynamics is embedded.} In other words, we want
\begin{align}
\hamt &= P_+ \ham P_+ + P_- \ham^\text{irr} P_-.
\end{align}
Using the basis \eqref{basis} any state in the relevant subspace $\hs$ can be written as:
\begin{align}
\label{ewavefunction}
\ket{\psit(t)} &= \sum_{n \in \fB_{N}} c_n(t) \ket{n_+} \, .
\end{align}
Our goal is to find a unitary transformation $\ut$ (see Eq.\eqref{basis}) and a Hamiltonian $\hamt$ without the many-body interaction $\op^\chi$, such that the coefficients in \eqref{ewavefunction} exactly follow those in \eqref{wavefunction}.
In other words the (N+1)-qubit basis state $\ket{n_+}$ will ``stand for" or ``encode" the N-qubit basis state $\ket{n}$.  

We write the Hamiltonian in the extended Hilbert space as:
\begin{align}
\hamt &= \sum_{\At \in \fQ_{N+1}} \hamt^\At \, .
\end{align}
Each $\hamt^\At$ will be chosen to simulate the dynamics that $\ham^A$ of \eqref{originalH} generates in the original system. 

The Schr\"odinger equation for the \target system in the chosen basis can be written as:
\begin{align}
\label{Schrodinger}
i \hbar \dot{c}_k &= \sum_n H_{kn} c_n \, ,
\end{align}
where $H_{kn} \equiv \bra{k} \ham \ket{n}$.
Similarly the Schr\"odinger equation for the \physical system can be written as:
\begin{align}
\label{eSchrodinger}
i \hbar \dot{c}_{k} &= \sum_{n} \hamt_{k_+ n_+} c_{n} \, ,
\end{align}
where $\hamt_{k_+ n_+}\equiv\bra{k_+} \hamt \ket{n_+}$.
As stated earlier, we want the physical Hilbert space to split into two decoupled Hilbert spaces under the dynamics imposed by the physical Hamiltonian $\hamt$. 
This condition can be expressed as
\begin{align}
\nonumber
\bra{k_+} \hamt \ket{n_-} &= 0\, , \\
\bra{k_-} \hamt \ket{n_+} &= 0 \, . 
\end{align} 
Using Eq.~\eqref{basis} these conditions can be rewritten as
\begin{align}
\nonumber
\bra{0 k} \Kt \ket{0 n} - \bra{0 k} \Kt \ket{1 n} + \bra{1 k} \Kt \ket{0 n} - \bra{1 k} \Kt \ket{1 n} &= 0 , \\
\bra{0 k} \Kt \ket{0 n} + \bra{0 k} \Kt \ket{1 n} - \bra{1 k} \Kt \ket{0 n} - \bra{1 k} \Kt \ket{1 n} &= 0 ,
\end{align}
where $\Kt$ is defined as
\begin{align}
\label{K}
\Kt &= \ut^\dagger \hamt \ut\, .
\end{align}

Adding and subtracting these two lines we obtain a simple expression for the conditions required for the two subspaces to be decoupled:
\begin{align}
\nonumber
\bra{0 k} \Kt \ket{0n} &= \bra{1k} \Kt \ket{1 n} \, , \\
\label{sscond}
\bra{0 k} \Kt \ket{1n} &= \bra{1k} \Kt \ket{0 n}\, .
\end{align}
The first line implies that $\Kt$ can not act on the ancilla qubit with a $\sigma^z$ operator, whereas the second line rules out the $\sigma^y$ operator. Thus
\begin{align}
\label{form1}
\Kt &= \id \otimes K^* + \sigma^x \otimes K^\chi \, .
\end{align}
The reason for this choice of superscripts will become clear shortly. 
Next, we determine the conditions for the dynamics in $\hs$ to simulate the dynamics of interest due to the \target Hamiltonian $\ham$. 
This is simply read from Eqs.(\ref{Schrodinger}, \ref{eSchrodinger}):
\begin{align}
&\bra{k_+} \hamt \ket{n_+} = \bra{k} \ham \ket{n}\, .
\end{align}
Using the conditions \eqref{sscond} this expression simplifies to
\begin{align}
\label{dyncond}
\bra{0k}\Kt \ket{0n} + \bra{0k} \Kt \ket{1n} &= \bra{k} \ham \ket{n} \, ,
\end{align}
which can be further simplified by using \eqref{form1} and \eqref{originalH}:
\begin{align}
\bra{k} K^* \ket{n} + \bra{k} K^\chi \ket{n} = \bra{k} \Hloc \ket{n} + \bra{k} H^\chi \ket{n}\, .
\end{align}
Since the \target Hamiltonian \eqref{originalH} is split into two parts it is natural to make the assignment $K^* = \Hloc$ and $K^\chi = H^\chi$, which leads to
\begin{align}
\Kt &= \ut^\dagger \hamt \ut = \id \otimes \Hloc + \sigma^x \otimes H^\chi \, ,\\
\hamt &= \ut \Kt \ut^\dagger = \ut \left( \id \otimes \Hloc + \sigma^x \otimes H^\chi \right) \ut^\dagger \equiv \hamt^* + \hamt^{\tilde{\chi}}\, .
\label{Ht}
\end{align}
Using Eq.~\eqref{Ht} we can also calculate the matrix elements of the Hamiltonian in the \junk subspace $\hsjunk$
\begin{align}
\label{Hwrong}
\bra{k_-} \hamt \ket{n_-} &= \bra{k} (\Hloc - H^\chi ) \ket{n} \equiv \bra{k} \Hwrong \ket{n} \, ,
\end{align} 
where we defined the ``irrelevant" Hamiltonian $\Hwrong\equiv \Hloc-H^\chi$ as the original problem Hamiltonian with the sign of the nonlocal term reversed. 
Thus, the dynamics in the \junk subspace $\hsjunk$ is closely related to the desired dynamics in computational subspace $\hs$. 
This also shows that unlike perturbative gadgets, here $\hs$ is in general not the low energy subspace. 
Generically, the energy levels of both subspaces are intermingled.   
Let us assume that the eigenstates and eigenvalues of the \target and \junk Hamiltonian are given by $(\ket{\psi^n_+},E^n_+)$ and $(\ket{\psi^n_-},E^n_-)$ respectively. 
Then it is straightforward to show that the eigenfunctions and eigenvalues of the \physical Hamiltonian are given by
\begin{align}
\ket{\psit_\pm^n} &= \ut \left( \ket{\pm} \otimes \ket{\psi_\pm^n}\right)\, , \quad \wt{E}^n_\pm = E^n_\pm\, .
\end{align}

Our strategy is to ``transfer" the nonlocality associated with the \target Hamiltonian to the unitary transformation $\ut$ such that the \physical Hamiltonian is free of that many-body interaction. 
By demanding that the nonlocal term $\op^\chi$ be absent from $\hamt$, a nonlocality is introduced to $\ut$ as a compensation.
Since the unitary transformation corresponds to a simple change of basis, the nonlocality therein does not present as serious a challenge as a many-body interaction in the Hamiltonian.
We will comment on this further in Sec.~\ref{sec:unitary}.

In particular we wish to have $\hamt^{\tilde{\chi}}$ to be a $r$-body interaction with small $r$.   
There are different choices with different trade-offs, and below we treat them separately. 

\subsection{Case 1: $\hamt^{\tilde{\chi}}$ is $1$-local}
\label{sec:1local}

It is possible to choose $\ut$ such that the term in $\hamt$ corresponding to the nonlocal term $\ham^\chi$  is only $1$-local:
\begin{align}
\label{U1}
\ut &= \ket{0}\bra{0} \otimes \id + \ket{1}\bra{1}\otimes \op^\chi\, . 
\end{align}
It is straightforward to verify that this operator is both unitary and Hermitian.
From \eqref{Ht} we get:
\begin{align}
\nonumber
\hamt &= \ket{0}\bra{0}\otimes \Hloc + \ket{1}\bra{1} \otimes \op^\chi \Hloc \op^\chi \\
\label{Httemp}
&\quad+ h_\chi \left(\ket{0}\bra{1} + \ket{1}\bra{0} \right)\otimes \id \, .
\end{align}
We next split the Hamiltonian $\Hloc$ into two parts based on the commutation properties of the terms with the nonlocal term $\op^\chi$.
\begin{align}
\Hloc &= \sum_{A\ne \chi} H^A = \Hloc_\comm + \Hloc_\anti \, , \\
0 &= [\Hloc_\comm , \op^\chi] \, , \\
0 &= \{ \Hloc_\anti, \op^\chi \} \, .
\end{align}
With this definition \eqref{Httemp} becomes
\begin{align}
\label{H1local}
\hamt &= \id \otimes \Hloc_\comm + \sigma^z \otimes \Hloc_\anti + h_\chi \left( \sigma^x \otimes \id \right) \, .
\end{align}
Note the price that had to be paid in order to eliminate the nonlocal term $\ham^\chi$: the locality of some other terms $\Hloc_\anti$ had to be increased by one. 
The locality of other terms $\Hloc_\comm$ which commute with the nonlocal term stayed the same, except for the nonlocal term $\ham^\chi$ itself, which we reduced to $1$-local.

Intuitively, one can think of the ancilla qubit as a sophisticated bookkeeping tool. 
The nonlocal term $\op^\chi$ acting on many system qubits is replaced with a simple spin flipping term $\sigma^x$ acting on the ancilla qubit only. 
Thus the state of the ancilla qubit ``keeps track" of the intended applications of the nonlocal term to the system during the evolution. 
The quantum nature of this bookkeeping is manifest in the modification of the rest of the Hamiltonian according to commutation rules.  
A similar technique has been developed independently by M. R. Geller~\cite{Geller2015arXiv} in the context of the single-excitation subspace method whereby each ancilla controls the application of an arbitrary $n\times n$ unitary to the data. 

\subsection{Case 2: $\hamt^{\tilde{\chi}}$ is $r$-local}
\label{sec:manylocal}

In this section we eliminate the nonlocal term in favor of an $r$-local term. At first sight this seems counterproductive but we will point out some cases in which this strategy proves to be advantageous. 
Let us consider a decomposition of the nonlocal operator of the form:
\begin{align}
\op^\chi &= \op^{\chi'} \op^{\chi''} = \op^{\chi''} \op^{\chi'} \, .
\end{align}
As an example:
\begin{align}
\nonumber
\op^\chi &= \sigma^x_1\otimes \sigma_2^y \otimes \sigma_3^z \otimes \id \otimes \sigma_4^y \, , \\
\nonumber
\op^{\chi'} &= \sigma^x_1\otimes \sigma_2^y \otimes \id \otimes \id \otimes \id \, , \\
\op^{\chi''} &= \id \otimes \id \otimes \sigma_3^z \otimes \id \otimes \sigma_4^y \, .
\end{align}
Next we define the basis transformation as:
\begin{align}
\ut &= \ket{0}\bra{0} \otimes \id + \ket{1}\bra{1}\otimes \op^{\chi'} \, .
\end{align}
Substituting this into Eq.~\eqref{Ht} we get the Hamiltonian in the extended Hilbert space:
\begin{align}
\label{Htildemany}
\hamt &= \id \otimes \Hloc_\comm + \sigma^z \otimes \Hloc_\anti + h_\chi \left( \sigma^x \otimes \op^{\chi''} \right) \, , \\
0 &= [\Hloc_\comm , \op^{\chi'}] \, , \\
0 &= \{ \Hloc_\anti, \op^{\chi'} \} \, .
\end{align}
Note that the splitting $\Hloc = \Hloc_\comm + \Hloc_\anti$ depends on the decomposition $\op^\chi = \op^{\chi'} \op^{\chi''}$. 
It is preferable for $\Hloc_\anti$ to have as few terms as possible and those terms to be as local as possible. 
Moreover, $\op^{\chi''}$ should be as local as possible. 
In general, these demands can be contradictory and there is no unique way to optimize the choice of $\op^{\chi'}$ independent of context. 
However, it should be clear that there can be an advantage to using the method of this section as opposed to the previous one, and we will provide an example related to adiabatic quantum algorithms in Sec.~\ref{sec:manyspecial}.

\section{Application to Adiabatic Quantum Algorithms}
\label{sec:AQCapplication}

The discussion in this section will be restricted to combinatorial optimization problems. 
The problem is embedded in a ``problem Hamiltonian" which is diagonal in the computational basis
\begin{align}
\ham^P = \sum_n E_n \ket{n} \bra{n} \, .
\end{align}
Being diagonal in the computational basis the problem Hamiltonian can be written in terms of the action of $\sigma^z$ and $\id$ operators only. Thus we can expand:
\begin{align}
\label{Hpexpand}
\ham^P = h  \id + \sum_i h_i \sigma_i^z + \sum_{ij} h_{ij} \sigma_i^z \sigma_j^z + \sum_{ijk} h_{ijk} \sigma_i^z \sigma_j^z \sigma_k^z + \dots 
\end{align}
One then considers a time-dependent Hamiltonian which extrapolates between an ``initial Hamiltonian" $\ham^0$ (also called ``driver Hamiltonian") and the problem Hamiltonian according to a predetermined schedule 
\begin{align}
\label{protocol}
\ham(t) = f(t) \ham^0 + g(t) \ham^P \, ,
\end{align}
where $f(0)=g(\tau)=1$ and $f(\tau)=g(0)=0$ and $\tau$ is the duration of computation. 

The adiabatic algorithm works as follows~\cite{Farhi2002arXiv}: 
the initial state is prepared to be the ground state of $\ham^0$. 
If the duration of computation is long enough, the adiabatic theorem guarantees that the system will stay arbitrarily close to the instantaneous ground state at all times. 
The ground state of the final Hamiltonian encodes the solution of the optimization problem and can be read via a measurement in the computational basis. 

The initial Hamiltonian is usually chosen to be:
\begin{align}
\label{H0std}
\stdham^0 = \sum_{i=1}^N B_i \, \sigma_i^x \, .
\end{align}
We will refer to this as the ``standard initial Hamiltonian". 
In some of the examples below we will also consider modified initial Hamiltonians with many-body interactions which are made out of tensor product of multiple $\sigma^x$ operators. 
 
\subsection{$\sigma^x$ -- only interactions}
\label{sec:xonly}

Consider the nonstandard initial Hamiltonian
\begin{align}
\ham^0 = \stdham^0 + B_0 \left( \sigma_1^{\beta_1} \otimes \sigma_2^{\beta_2} \otimes \dots \otimes \sigma_N^{\beta_N} \right) \equiv \stdham^0 + \ham^\chi \, ,
\end{align}
where $\beta_i\in \{0,x\}$. 
We follow the recipe of Sec.~\eqref{sec:1local} to simulate the last term with a $1$-local term . 
We first need to determine which terms of the Hamiltonian do not commute with $\ham^\chi$.
It is clear that $\stdham$ commutes with $H^\chi$. 
For the problem Hamiltonian we refer to Eq.\eqref{Hpexpand}. 
Whenever a term has an even number of qubits that get acted nontrivially by both $\ham^\chi$ and the element of $\ham^P$ in question, those two operators commute. 
For odd number of common qubits the two operators anti-commute.
For example if
\begin{align}
\ham^\chi &\propto \sigma_1^x \otimes \sigma_2^x \otimes \id_3\otimes \sigma_4^x \otimes \id_5 
\end{align}
then $h_3 \sigma_3^z$, $h_{12} \sigma_1^z \sigma_2^z$ and $h_{145} \sigma_1^z \sigma_4^z \sigma_5^z$ commute with $\ham^\chi$ but $h_1 \sigma_1^z$, $h_{23} \sigma_2^z \sigma_3^z$ and $h_{123} \sigma_1^z \sigma_2^z\sigma_3^z$ anti-commute 
We group these terms together and rewrite the problem Hamiltonian as:
\begin{align}
\ham^P &= \ham^P_\comm + \ham^P_\anti \, ,
\end{align}
where $\ham^P_\comm$ consists of terms that commute with $\ham^\chi$ and $\ham^P_\anti$ those that  anti-commute with it.
As a result the \physical Hamiltonian is given by
\begin{align}
\nonumber
\hamt &= f(t) \left[ B_0 \left( \sigma^x \otimes \id \right) + \stdham^0 \right] \\ 
&\hspace{15mm} + g(t) \left[ \id\otimes \ham^P_\comm + \sigma^z \otimes \ham^P_\anti \right] \\
\label{AQCxxx}
&\equiv f(t) \wt{\ham}^0_\text{std} + g(t) \wt{\ham}^P \, .
\end{align}
Notice that by treating the ancilla qubit as the $0$'th qubit $\hamt$ takes exactly the same form as $\ham$, the only difference being the number of qubits. 
This example shows us that a nonlocal term consisting of $\sigma^x$ operators only can be incorporated into the initial Hamiltonian of an adiabatic quantum algorithm at the price of increasing the nonlocality of some of the terms in the problem Hamiltonian by one but without changing the form of the Hamiltonian.

\subsection{$\sigma^z$ -- only interactions}
\label{sec:zonly} 

In this section we consider the standard adiabatic algorithm with $\stdham^0$ given by Eq.\eqref{H0std}. 
The nonlocal interaction we would like to eliminate is a product of $\sigma^z$ operators only, thus it is part of the problem Hamiltonian $\ham_P$: 
\begin{align}
\ham^\chi \propto \sigma_1^{\beta_1} \otimes \sigma_2^{\beta_2} \otimes \dots \otimes \sigma_N^{\beta_N}\, ,
\end{align}
where $\beta_i \in \{0,z\}$. As a result $\ham^\chi$ commutes with $\ham^P$ but not with all the terms in $\stdham^0$. In fact, for all $i$ such that $\beta_i = z$, the corresponding term $B_i \sigma_i^x$ does anti-commute with $\ham^\chi$. If we group these terms in a manner similar to the previous section we can rewrite the Hamiltonian as:
\begin{align}
\ham(t) &= f(t) \left( \ham^0_\comm + \ham^0_\anti \right) + g(t) \ham^P \, .
\end{align}
A derivation analogous to the previous section results in the following Hamiltonian for the \physical Hilbert space
\begin{align}
\hamt &= f(t) \left[ \id \otimes \ham^0_\comm + \sigma^z \otimes \ham^0_\anti \right] + g(t) \left( \id \otimes \ham_P \right) \, .
\end{align}
In contrast to the previous section this Hamiltonian is not of the same form as the original Hamiltonian.
The difference is the $\sigma^z \otimes \sigma^x$ type interactions in the second term above, i.e., $\sigma^z \otimes \ham^0_\anti$. 
On the other hand, the locality of the physical Hamiltonian is the same as the locality of $H^*$, i.e., the computational Hamiltonian without the many-body interaction term $H^\chi$.

\subsection{The spin glass problem}

In this section we apply the technique described in this note to various adiabatic quantum algorithms for finding the ground state of the spin glass problem. 
The spin glass problem Hamiltonian is given by
\begin{align}
\label{spinglass}
\ham^P = \sum_{i=1}^N h_i \sigma_i^z + \sum_{i>j=1}^N h_{ij} \sigma_i^z \sigma_j^z \, .
\end{align}

\subsubsection{Flip-All Term}
\label{sec:flipall}

A nonlocal interaction of the form given in Sec.~\eqref{sec:xonly} that leads to a particularly simple physical Hamiltonian is 
\begin{align}
\label{flipall}
\ham^\chi = B_0 \left( \sigma_1^x\otimes \sigma_2^x\otimes \dots \otimes \sigma_N^x \right) \, .
\end{align}
This is an $N$-body interaction, involving all \target qubits. 
When acting on a basis state it flips the orientation of all the spins, hence the title of this section. 
Our goal is to simulate the dynamics due to 
\begin{align}
\ham(t) &= f(t) \left( \stdham^0 + \ham^\chi \right) + g(t) \ham^P
\end{align}
without actually implementing $\ham^\chi$ in the \physical Hamiltonian.
It is clear that all the $2$-local terms in the problem Hamiltonian \eqref{spinglass} commute with $\ham^\chi$ and so does the initial Hamiltonian of the standard algorithm. 
On the other hand, all the $1$-local terms in the problem Hamiltonian anti-commute with $\ham^\chi$. 
This is a special case of Sec.~\eqref{sec:xonly} and we can directly read off the \physical Hamiltonian:
\begin{align}
\label{flipallsolution}
\nonumber
\hamt(t) &= f(t) \sum_{i=0}^N B_i \, \sigma_i^x + g(t) \sum_{i>j=0}^{N} \tilde{h}_{ij} \left( \sigma_i^z \otimes \sigma_j^z \right) \\
&= f(t) \wt{\ham}_\text{std}^0 +g(t) \hamt^P \, , \\
\tilde{h}_{ij} &= h_{ij} \qquad \text{for } i,j \ne 0 \, , \\
\tilde{h}_{0j} &= h_j \qquad \text{for } j = 1,2,\dots,N \, .
\end{align}
This Hamiltonian has the form of a standard adiabatic algorithm to solve a spin glass problem with vanishing local fields.
The ancilla qubit is coupled to all qubits that are acted on by local fields in the original problem Hamiltonian. 

This means that for every spin glass problem of $N$ qubits solved using the algorithm with the modified initial Hamiltonian $\ham^0_\text{std}+\ham^\chi$, there is an equivalent $N+1$-qubit problem which can be solved with the standard initial Hamiltonian, such that the success rate of both algorithms is identical.  

\subsubsection{A case for Sec.~\ref{sec:manylocal}}
\label{sec:manyspecial}

As mentioned before, there are problems for which the approach of Sec.~\ref{sec:manylocal} works better than that of Sec.~\ref{sec:1local}. 
Consider the spin glass problem with the standard initial Hamiltonian modified by adding the following term
\begin{align}
\label{flipallbut}
\ham^\chi = B_0 \left( \sigma_2^x\otimes \sigma_3^x\otimes \dots \otimes \sigma_N^x \right) \, .
\end{align}
This differs from Eq.\eqref{flipall} by the absence of $\sigma_1^x$ and thus represents a $(N-1)$-body interaction. 
If we follow the approach of Sec.~\ref{sec:1local} we 
get (again treating the ancilla as the 0'th qubit):
\begin{align}
\nonumber
\hamt(t) &= f(t) \left(\sum_{i=0}^N B_0 \, \sigma_i^x \right) + g(t) \Bigg( h_1 \sigma_1^z + \sum_{i=2}^N h_i \sigma_0^z \sigma_i^z \\& +  \sum_{i=2}^N J_{1i}\, \sigma_0^z \sigma_1^z \sigma_i^z + \sum_{i>j=2}^N J_{ij} \sigma_i^z \sigma_j^z \Bigg) \, .
\end{align}
This is a $3$-local Hamiltonian if at least one of the $J_{1i}\ne 0$ for $i=2,\dots,N$. 

If, on the other hand, we follow the approach of Sec.~\ref{sec:manylocal} with the choice:
\begin{align}
\op^{\chi'} &= \sigma_1^x \otimes \sigma_2^x \otimes \dots \otimes \sigma_N^x \, ,\\
\op^{\chi''} &= \sigma_1^x \otimes \id_2 \otimes \dots \otimes \id_N \, .
\end{align}
The resulting Hamiltonian is:
\begin{align}
\nonumber
\hamt(t) &= f(t) \left( B_0 \left(\sigma_0^x \otimes \sigma_1^x \right) + \sum_{i=1}^N B_i \, \sigma_i^x \right) \\
\label{flipallbutsolution}
&\qquad+ g(t) \left( \sum_{i>j=0}^{N} \tilde{h}_{ij} \left(\sigma_i^z \otimes \sigma_j^z\right) \right) \\
\nonumber
&= f(t) \left( B_0 \left( \sigma_0^x \otimes \sigma_1^x \right) + \id \otimes \stdham \right) +g(t) \hamt_P \, , \\
\tilde{h}_{ij} &= h_{ij} \qquad \text{for } i,j \ne 0 \, , \\
\tilde{h}_{0i} &= h_i \qquad \text{for } i = 1,2,\dots,N \, .
\end{align}
This differs from Eq.\eqref{flipallsolution} only by the replacement $B_0\, \sigma_0^x \rightarrow B_0 ( \sigma_0^x \otimes \sigma_1^x)$. 
Note that \eqref{flipallbutsolution} is only $2$-local. 
This example demonstrates how sometimes it can pay off to simulate the many-body interaction with a $(r>1)$-body interaction instead of a simple field, i.e. a $1$-local term.

\section{Multiple Nonlocal Terms}
\label{sec:multiple}

In the previous section we discussed how to eliminate a nonlocal term in a Hamiltonian. 
If we wish to eliminate multiple nonlocal terms, the technique can be applied repeatedly. 
However, there is a problem with this strategy, because at each step the degree of locality of some terms in the Hamiltonian increases by one. 
Consider a Hamiltonian with $\ntot$ terms $\nloc$ of which are $2$-local and $\nnloc$ of them are more than $2$-local such that $\ntot = \nloc + \nnloc$. 
In eliminating the $\nnloc$ terms using the approach described in this manuscript repeatedly, we possibly raise the degree of nonlocality in other terms. 
In the worst case scenario, we may end up with terms which have a degree of nonlocality $2+\nnloc$ .

A better strategy might be to use perturbative Hamiltonian gadgets at the end of each round to reduce the degree of locality of those terms that have been raised from 2 to $3$-local. 
Using this hybrid method the use of the perturbative gadgets is restricted to 3-body interactions only, where they work best.  

However there is a trade-off. 
For each nonlocal term eliminated many $3$-local terms are created. 
There are two possible problems with this. 
First, the number of ancilla qubits necessary does not only depend on the degree of nonlocality but possibly also on the number of total qubits in the system. 
Thus it might scale very badly.
The second problem is related to the perturbative nature of gadgets. If errors due to different gadgets accumulate coherently we might run into trouble.  

\section{Implementation}
\label{sec:implementation}
\subsection{State Preparation}
\label{sec:statepreparation}
\begin{figure*}[ht]
	\centering
	\includegraphics[width=0.85\linewidth]{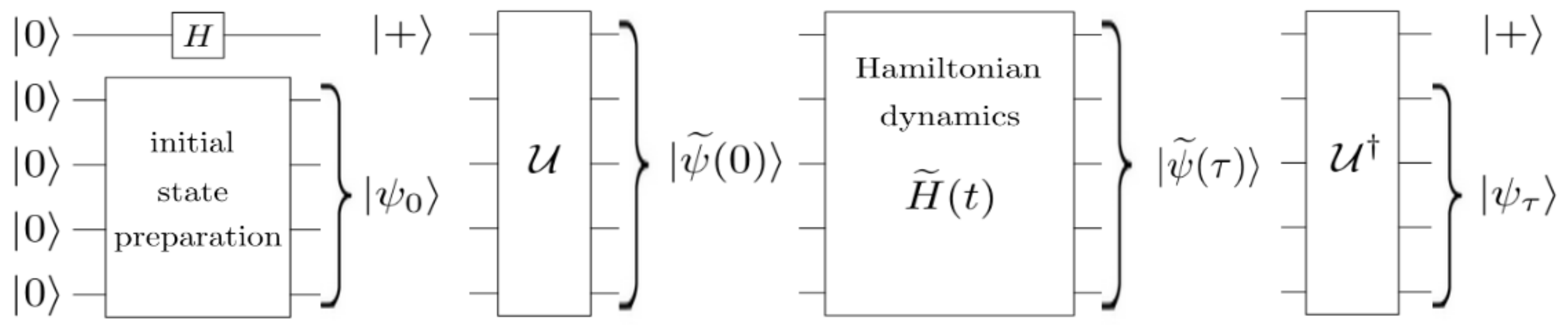}
	\caption{A schematic description of the state preparation protocol described in detail in Sec.~\ref{sec:statepreparation}. The top qubit is the ancilla and the rest are system qubits. $H$ acting on the ancilla stands for the Hadamard gate, not to be confused with the Hamiltonian. }
	\label{fig:fullprep}
\end{figure*}

In previous sections we showed how to construct an N+1 qubit system such that the dynamics in a subspace is mathematically identical to an N qubit system. 
However, we did not address the question of how to use this mapping in an operational sense. 
To this end let us consider the state preparation protocol. 
A schematic description is given in Fig.~\ref{fig:fullprep}. 

In state preparation, the goal is to prepare a target quantum state $\ket{\psi_\tau}$. 
In some cases the target state is known ahead of time but in others it is defined as the output of a certain process. 
Here we are interested in the latter. 
More specifically, we are interested in the evolution of a given initial state $\ket{\psi_0}$ of N qubits for a time $\tau$ according to Hamiltonian dynamics with time-dependent Hamiltonian $\ham(t)$. 
At the final time $\tau$ the desired N-qubit state $\ket{\psi_\tau}$ is obtained. 
State preparation can serve as a subroutine of a larger computation or as part of a larger simulation.

Many-body interactions can be used to increase the success rate of adiabatic state preparation by allowing one to explore a larger space of Hamiltonian paths from the initial to the final Hamiltonian. 
There is preliminary numerical evidence for this claim for the spin glass problem and the initial Hamiltonian studied in Sec.~\ref{sec:flipall}~\cite{inprep}. 

We assume that we have the ability to prepare the N-qubit initial state $\ket{\psi_0}$. 
In order to implement the technique described in previous sections, we need to encode this state in the N+1 qubit Hilbert space using \eqref{ewavefunction}. 
First the ancilla qubit is initialized to the superposition state $\ket{+}$.  
Then one implements the unitary operator $\ut$ on the combined system (the details are discussed in the next section)
\begin{align}
\nonumber
\ket{\psit(0)} &= \ut \left(\ket{+} \otimes \ket{\psi_0} \right) \\ 
\label{initialstate}
&= \frac{\ket{0}\otimes \ket{\psi_0} + \ket{1} \otimes \op^\chi\ket{\psi_0}}{\sqrt{2}}\, .
\end{align} 
Next, the initial state $\ket{\psit(0)}$ is evolved with the Hamiltonian $\hamt$ of \eqref{H1local}. After a time $\tau$ the following state is obtained:
\begin{align}
\label{finalstate}
\ket{\psit(\tau)} &=  \frac{\ket{0}\otimes \ket{\psi_\tau} + \ket{1} \otimes \op^\chi\ket{\psi_\tau}}{\sqrt{2}}\, ,\\
&=\ut \left(\ket{+} \otimes \ket{\psi_\tau} \right) \, . 
\end{align}
One then applies the inverse unitary transformation $\ut^\dagger = \ut$ to obtain the state: 
\begin{align}
\ut^\dagger \ket{\psit(\tau)} = \ket{+} \otimes \ket{\psi_\tau} \, .
\end{align} 
This completes the state preparation protocol. 
At this point one can ignore the ancilla qubit altogether, and the rest of the $N$ qubits are in the target state. 

The ancilla qubit does not need to be measured but a measurement in the $\ket{\pm}$ basis can help detect some errors that may have occurred. 
If the ancilla is found in the $\ket{-}$ state, it is an indication that the N+1 qubit system has been knocked out of the relevant subspace $\hs$ into $\hsjunk$ and the state preparation is not to be trusted.

\subsection{Implementing the Unitary $\ut$}
\label{sec:unitary}

Note that this unitary is a simple product of 2-qubit controlled-U gates of the form
\begin{align}
	\nonumber
	\ut &= \ket{0}\bra{0} \otimes \id + \ket{1}\bra{1} \otimes \op^\chi \\
\label{Uhow}
	&= \prod_{i=1}^N  \bigg( \ket{0}\bra{0} \otimes \id + \ket{1}\bra{1} \otimes \sigma_i^{\chi_i} \bigg) \, ,
\end{align}
where in the second line we abused the notation by letting $\ket{1}\bra{1} \otimes \sigma_i^{\chi_i} \equiv \ket{1}\bra{1} \otimes \id_1 \otimes \cdots \otimes \sigma_i^{\chi_i} \otimes \cdots  \otimes\id_N$. 
The product is over those $i$ such that $\chi_i \ne 0$ only, because all other terms are the identity. 
Thus for a $k$-body interaction term $\op^\chi$, $\ut$ can be implemented by repeatedly applying $k$ 2-qubit controlled-U gates (more specifically controlled-X,Y,Z gates)~\cite{Nielsen2010Book}. 
This can be easily achieved in a gate based (digital) quantum computing model. 

The problem of realizing many-body interactions we are addressing in this paper is more pertinent to the adiabatic quantum computing model. 
Unlike the gate model, AQC is not built upon the concept of gates.  
Yet both paradigms of quantum computing have been shown to be polynomially equivalent in terms of complexity theory~\cite{Aharonov2008SIAM}.   
Recently, Hen~\cite{Hen2015PRA} made a connection between these two paradigms by showing that a universal set of quantum gates can be realized within the AQC framework via controlled adiabatic evolutions. 

Hence the technique developed in this paper can be applied to AQC by using Hen's quantum adiabatic algorithms for gates as subroutines for implementing the unitary $\ut$ at the beginning and at the end of the protocol.  
The original method due to Hen requires a single auxiliary qubit. The runtime scales linearly with $k$ and is independent of $N$, the total number of qubits. 
Shortcuts to adiabaticity can be used to realize these gates with unit probability at finite time~\cite{Santos2015arXiv}.

\subsection{A special case}
\label{sec:special}

Let us now consider the case when the initial state is invariant under the action of the many-body interaction term, i.e., 
\begin{align}
\op^\chi \ket{\psi_0} = \ket{\psi_0}.   
\end{align}
This implies that $ \ut \left(\ket{+} \otimes \ket{\psi_0}\right) = \ket{+} \otimes \ket{\psi_0}$. 
Thus for this case the first application of the unitary transform prior to the Hamiltonian evolution is not necessary. 
\begin{figure*}[ht]
	\centering
	\includegraphics[width=0.70\linewidth]{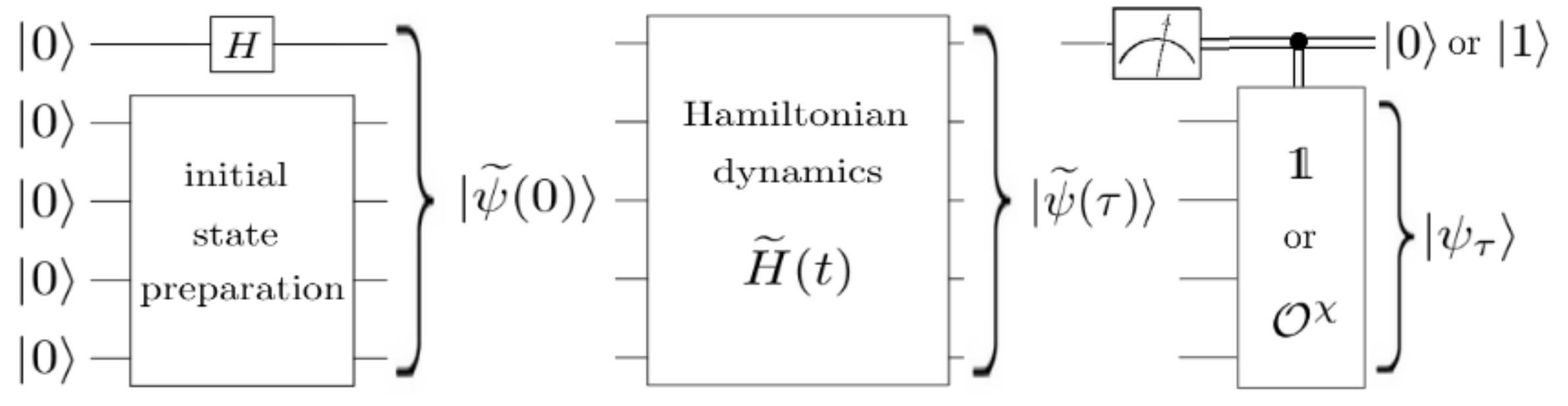}
	\caption{A schematic description of the state preparation protocol applicable only to the special case described in detail in Sec.~\ref{sec:special}. The top qubit is the ancilla and the rest are system qubits. $H$ acting on the ancilla stands for the Hadamard gate, not to be confused with the Hamiltonian. After the Hamiltonian dynamics, the ancilla qubit is measured and conditional on the outcome the system qubits are acted on by a different unitary.}
	\label{fig:shortcircuit}
\end{figure*}

A schematic description of the state preparation protocol applicable to this case is given in Fig.~\ref{fig:shortcircuit}. 
The state after the Hamiltonian evolution is given by \eqref{finalstate}. 
In this state the system and ancilla are entangled. 
In the previous section we suggested applying the inverse unitary to disentangle them. 
Another option is to measure the ancilla qubit first. 
If the ancilla is found to be in the $\ket{0}$ state the system qubits are already in the desired target state $\ket{\psi_\tau}$. 
If the ancilla is measured to be in the $\ket{1}$ state the system is in the state $\op^\chi \ket{\psi_\tau}$. 
This state can be transformed into the target state by acting on it with the unitary operator $\op^\chi$ again. 
The advantage of this approach  is that $\op^\chi$ can be realized as $k$ successive single qubit unitaries.
\begin{align}
\label{Ophow}
 \op^\chi = \prod_{i}  \sigma_i^{\chi_i}\, ,
\end{align}
where the product is over the indices with nonvanishing $\chi_i$. 
This is to be contrasted with $k$ successive two qubit unitaries necessary to implement $\ut$ shown in \eqref{Uhow}. 
Since two qubit unitaries are significantly more difficult to implement experimentally than single qubit unitaries the method of this section can provide great simplification whenever applicable.  

Adiabatic optimization algorithms, for which both the initial Hamiltonian and the many-body interaction term $\op^\chi$ are made up of $\sigma^x$ operators only, falls under this category. 
The initial state can be prepared simply by applying strong local fields to all N+1 qubits as in \eqref{AQCxxx}
Moreover, for the purpose of finding the answer to the optimization problem one does not even need to prepare the state $\ket{\psi_\tau}$ itself. 
One can perform the final step described in \eqref{Ophow} {\it on paper}. 
If the ancilla is found to be in state $\ket{0}$ and the rest of the qubits in a state $\ket{n}$, the outcome of the computation is simply given by $n$. 
If, on the other hand, the ancilla is found to be in state $\ket{1}$ and the measurement of the system yields $\ket{n}$, the outcome of the computation is interpreted as $\bar{n}$, such that $\op^\chi \ket{n} = \ket{\bar{n}}$. 
Since $\op^\chi$ is assumed to be a tensor product of $\sigma^x$ operators only, such an $\bar{n}$ always exists.

\section{Conclusion}
\label{sec:conclusion}

In this paper we presented an embedding of an N-qubit Hamiltonian with many-body interactions, into a subspace of an (N+1)-qubit Hamiltonian. 
In the simplest case, a many-body interaction term of the \target Hamiltonian is replaced with a spin flip operator $\sigma_x$ acting only on the ancilla qubit. 
As a concrete application of our method, in Sec.~\ref{sec:implementation} we discussed a state preparation protocol in detail. 
Our technique is nonperturbative and for a class of problems can be used to reduce the degree of nonlocality of the Hamiltonian significantly. 
We have discussed how it can be implemented in a hybrid as well as a solely adiabatic quantum computer.   
 
The success of the adiabatic quantum algorithms rely heavily on one parameter $\Delta$ which stands for the minimal spectral gap of the time-dependent Hamiltonian $\ham(t)$ along the path from $\ham(0)$ to $\ham(\tau)$. 
Eq.\eqref{protocol} specifies one such path for each choice of the functions $f$ and $g$. 
By choosing these wisely, the success rate can be increased significantly~\cite{Roland2002PRA,Lidar2009JoMP}. 
However \eqref{protocol} does not represent the most general Hamiltonian path from $\ham(0)$ to $\ham(\tau)$. 
For instance an arbitrary Hamiltonian can be turned on and off during the time interval $[0,\tau]$, without effecting the endpoints. 
Evidence that adding a random local  Hamiltonian to the middle of the adiabatic path increases the success probability has been presented in~\cite{Crosson2014arXiv}. 
It is conceivable that general paths involving nonlocal Hamiltonians at intermediate times could result in further likelihood of success. 
Similarly, using nonlocal initial Hamiltonians might also improve success rate in some cases. 
The technique developed in this paper might provide the means to realize such Hamiltonian paths using only few-body interactions. 

The nonlocality is one aspect that can make a many-body interaction challenging to realize experimentally at a fundamental level. 
A more practical difficulty is to couple a qubit to many other qubits, even via 2-body interactions. 
While remedying the former, the technique developed in this paper might exacerbate the latter in a given application. 
In the worst case, the ancilla qubit might be required to interact with all of the system qubits (see Sec.~\ref{sec:flipall}). 
In most experimental setups it is not feasible to have a complete interaction graph between all qubits.
However, it might be possible to design experiments where a small percentage of nodes (qubits) have very high connectivity. 
In such setups Hamiltonians with a few highly nonlocal interactions can be realized using the technique developed in this paper. 
Thus our analysis suggests a new design criteria for experiments to the extent that such Hamiltonians are relevant and useful in particular applications.  

\vspace{-1mm}

\acknowledgments

We gratefully acknowledge financial support from the Lockheed Martin Corporation under contract U12001C.
We would like to thank Oren Raz, Stephen Jordan, Paul Hess and Kanupriya Sinha for useful discussion.

\bibliography{HamiltonianEmbedding} 

\end{document}